\documentclass{emulateapj}

\newcommand{\tmin}{{T_{\rm min}}}
\newcommand{\msun}{M_{\odot}}
\renewcommand\ion[2]{#1$\;${\small\rmfamily{#2}}\relax}

\begin{document}

\title{How To Light It Up: Simulating Ram-Pressure Stripped X-ray Bright Tails}
\author{Stephanie Tonnesen, Greg L. Bryan and Rena Chen}
\affil{Department of Astronomy, Columbia University, Pupin Physics Laboratories, New York, NY 10027}

\begin{abstract}
Some tails of ram-pressure stripped galaxies are detected in \ion{H}{I}, some in H$\alpha$, and some in X-ray (but never all three so far).  We use numerical simulations to probe the conditions for the production of X-ray bright tails, demonstrating that the primary requirement is a high pressure intracluster medium (ICM).  This is because the stripped tail is mostly in pressure equilibrium with the ICM, but mixing leaves it with densities and temperatures intermediate between the cold gas in the disk and the hot ICM.  Given a high enough ICM pressure, this mixed gas lies in the X-ray bright region of the phase diagram.  We compare the simulations to observations of the ram pressure stripped tail of ESO 137-001, showing excellent agreement in the total measured X-ray and H$\alpha$ emission and non-flaring morphology of the tail, and consistent \ion{H}{I} measurements.  Using these comparisons we constrain the level of mixing and efficiency of heat conduction in the intracluster medium (ICM).

\end{abstract}


\section{Introduction}

Ram pressure (and related processes) by the intracluster medium (ICM) can remove a galaxy's gas (Gunn \& Gott 1972).  This process has been observed in various stages; for example, Vollmer (2009) separates Virgo galaxies into pre-peak, peak, and post-peak ram pressure groups.  The amount of time that a galaxy has been stripped can be estimated using the length of the observable tail and the velocity of the galaxy (e.g. Oosterloo \& van Gorkom 2005; Sun et al. 2006).  This calculation is uncertain due to difficulties in determining the three dimensional galaxy velocity.  Another assumption implicit in this calculation is that the observed tail provides the true length of the stripped gas.  In fact, tails have been observed in \ion{H}{I}, H$\alpha$, and X-ray emission, although never all from the same tail (e.g. Oosterloo \& van Gorkom 2005; Koopmann et al. 2008; Kenney et al. 2008; Yoshida et al. 2002; Yoshida et al. 2004a,b; Sun et al. 2007; Sun et al. 2006; Sun et al. 2010; Machacek et al. 2006; Sun \& Vikhlinin 2005).  The lengths of tails observed in different wavelengths can be quite different; for example the \ion{H}{I} tail of NGC 4388 is nearly three times as long as the observed H$\alpha$ tail (Oosterloo \& van Gorkom; Yoshida et al. 2002).  Another method used to calculate the age of a tail is to use the estimated survival time of H$\alpha$, as in Gavazzi et al. (2001).  However, it is still not clear what dictates cloud survival or even what conditions are necessary to produce the various types of emission (H$\alpha$, X-ray, and \ion{H}{I}).  Can all three types of emission coexist?  What physical processes dominate the heating and mixing of stripped gas into the ICM?  These processes include: turbulent mixing, which can generate intermediate temperature and density gas at constant pressure; shock heating, which heats the ISM; radiative cooling, which can lead to recompression of heated gas, and heat conduction, which can evaporate small clouds.
 
In this work we focus on answering these questions by simulating gas stripping and comparing our simulated tail to a single observed stripped galaxy, ESO 137-001, which has been studied observationally in some detail.  ESO 137-001 is in A3627, which is the closest massive cluster (z=0.0163, $\sigma_{radial}$ = 925 km s$^{-1}$ and kT = 6 keV), similar to Coma and Perseus in mass and galaxy content (Sun et al. 2009 and references therein).  ESO 137-001 is a small (0.2L$_*$; Sun et al. 2006), blue emission-line galaxy (Woudt et al. 2004), that is $\sim$200 kpc from the center of the cluster in projection.  Because its radial velocity is close to the average velocity of A3627 (Woudt et al. 2004; Woudt et al. 2008), most of its motion is likely in the plane of the sky, and therefore the stripping process is seen edge-on.  Sun et al. (2006) found a $\sim$70 kpc X-ray tail pointing away from the cluster center using \textit{Chandra} and XMM-\textit{Newton} data.  Sun et al. (2007) then discovered a 40 kpc H$\alpha$ tail with over 30 emission-line regions extending through the length of the H$\alpha$ tail, and concluded that the emission-line regions are giant \ion{H}{II} regions.  In a recent follow-up paper, Sun et al. (2009) used deep \textit{Chandra} data and \textit{Gemini} spectra to characterize the X-ray tail and \ion{H}{II} regions in detail.  They found a narrower secondary X-ray tail with a similar length.  They also confirmed that 33 emission-line regions are \ion{H}{II} regions, with the furthest seven regions beyond the tidally-truncated halo of 15 kpc that is calculated in Sun et al. (2007) using simulations by Gnedin (2003).  In addition to these distinct \ion{H}{II} regions, they find diffuse H$\alpha$ emission. 

Vollmer et al. (2001) searched for \ion{H}{I} in A3627, and did not detect any \ion{H}{I} in or around ESO 137-001 with a limiting column density of 2 $\times$ 10$^{20}$ cm$^{-2}$ and a resolution of 15".  In fact, of the $\sim$80 galaxies identified by Woudt et al. (1998) in their search region, Vollmer et al. (2001) detected only 2 in \ion{H}{I}, finding that the \ion{H}{I} detection rate in A3627 is similar to that in Coma.  

Sivanandam et al. (2009) observed ESO 137-001 with IRAC and IRS on \textit{Spitzer}.  The IRS data extended to 20 kpc from the galaxy along the X-ray tail, and warm ($\sim$160 K) molecular Hydrogen was detected throughout the length of the observed region.  The observed region contains $\sim$2.5 $\times$ 10$^7$ M$_\odot$ warm H$_2$ gas.  They also identify star-forming regions using 8 $\mu$m data, which coincide with H$\alpha$ emitting regions.  

There has been a substantial amount of theoretical work investigating ram pressure stripping in general (e.g. Schulz \& Struck 2001; Quilis, Bower \& Moore 2000; Roediger \& Br\"uggen 2008, Kronberger et al. 2008; Kapferer et al. 2009) -- see Tonnesen \& Bryan (2009, 2010; hereafter TB09 and TB10) for a more detailed discussion.  There have also been simulations designed to predict or interpret observational characteristics of ram pressure stripped tails and the remaining disks (e.g. Vollmer et al. 2005, 2006, 2008), but  detailed, quantitative predictions of all three observational probes have been missing to date (\ion{H}{I}, diffuse H$\alpha$, and X-ray emission).

In our previous work (TB10), we ran a set of high resolution simulations (about 38 pc resolution, which is small enough to marginally resolve giant molecular clouds) to understand how a multiphase ISM could affect the survival and structure of ram pressure stripped gas.  We focused on how density fluctuations that are observed in the multiphase ISM of galaxies can affect gas tails.  Including radiative cooling allowed us to estimate the density of and emission from \ion{H}{I}, H$\alpha$, and X-ray gas separately.  We found that both the morphology and velocity structure of our tails agreed with observations of long gas tails (e.g. Oosterloo \& van Gorkom 2005).  Our simulations also resulted in observable amounts of \ion{H}{I} and H$\alpha$ emission.  However, the X-ray tail had a low surface brightness, which we attributed to the low pressure of the surrounding ICM.  In this paper we will use the same method as in TB10, but have chosen ICM parameters comparable to the ICM around ESO 137-001.  By focusing on the level of agreement between our simulations and the observations of ESO 137-001 we will be able to discuss the importance of physical mechanisms such as heat conduction, as well as predict the conditions under which \ion{H}{I}, H$\alpha$, and X-ray emission are produced in stripped tails.

The paper is structured as follows.  After a brief introduction to our methodology, we provide the characteristics of our simulations and our method of producing simulated observations (\S 2).  We then (\S 3) present our results, specifically focusing on the comparison with ESO 137-001.  In \S 4 we discuss the broader implications of our simulation, and discuss our choice of radiative cooling floor and resolution in \S 5.1-2.  Finally, we conclude in \S 6 with a summary of our results and predictions for observers.

\begin{table}
\begin{center}
\caption{Galaxy Stellar and Dark Matter Constants\label{tbl-const4}}
\begin{tabular}{c | c}
\tableline
Variable & Value\\
\tableline
M$_*$ & $1 \times 10^{11}$ M$_{\odot}$ \\
a$_*$ & 4 kpc \\
b$_*$ & 0.25 kpc\\
M$_{bulge}$ & $1 \times 10^{10}$ M$_{\odot}$ \\
r$_{bulge}$ & 0.4 kpc \\
r$_{DM}$ & 23 kpc \\
$\rho_{DM}$ & $3.8 \times 10^{-25}$ g cm$^{-3}$ \\
\tableline
\end{tabular}
\end{center}
\end{table}

\section{Methodology}
\label{sec-methodology}

We use the adaptive mesh refinement (AMR) code {\it Enzo} (Bryan 1999; Norman \& Bryan 1999; O'Shea et al. 2004).  Our simulated region is 311 kpc on a side with a root grid resolution of $128^3$ cells.   We allow an additional 6 levels of refinement, for a smallest cell size of 38 pc.  We refine our simulation based on the local gas mass, such that a cell was flagged for refinement whenever it contained more than about $2 \times 10^4$ $\msun$.  We found that this refined most of the galactic disk to 38 pc resolution by the time the wind hit; dense clumps in the wake were also refined to 38 pc resolution, while more diffuse components had lower resolution.

The simulation includes radiative cooling using the Sarazin \& White (1987) cooling curve extended to low temperatures as described in Tasker \& Bryan (2006).  To mimic effects that we do not model directly (such as turbulence on scales below the grid scale, UV heating, magnetic field support, or cosmic rays), we cut off the cooling curve at a minimum temperature $T_{\rm min}$ so that the cooling rate is zero below this temperature.  In the simulations described here we use either $\tmin = 8000$ K, or $\tmin = 300$ K.  Both allow gas to cool below the threshold for neutral Hydrogen formation.  In TB09, we found that the minimum temperature affected the range of masses and sizes of clouds forming in the disk, and the distribution of clouds throughout the disk, although the range of gas densities was very similar.  The $\tmin = 300$ K case resulted in a more fragmented disk whose small clouds took longer to strip.  Therefore the timescales for gas stripping differed, although the total amount of gas lost was similar.  In the simulations presented in this paper, we find that the cooling floor affects the amount of fragmentation in the disks, but not the mass-distribution of gas densities. The larger radii of the remaining gas disks in the two $\tmin = 300$ K runs are due to the survival of dense clouds in the outer disk (Figures \ref{fig:xrayproj}, \ref{fig:haproj}, and \ref{fig:hiproj}). 

In TB10, we found that the structure of the wake also depends somewhat on $\tmin$.  Since the publication of TB10, we found and corrected an error in our implementation of the cooling rate that resulted in a slight shift in the peak of the cooling curve around $10^4$ K, but did not significantly affect cooling at low and high temperatures (this affected only the $\tmin = 300$ K simulation).  Tests showed that this had only a small impact on the dynamics of the flow, and on the predicted \ion{H}{I} and X-ray measures, but did strongly affect the predicted H$\alpha$ emission, which is extremely sensitive to the gas temperature around $10^4$ K.  This explains the enhanced H$\alpha$ emission in this paper for the $\tmin = 300$ K run as compared to TB10.

\begin{table}
\begin{center}
\caption{Gas Disk Constants\label{tbl-gconst4}}
\begin{tabular}{c | c }
\tableline
Variable & Value \\
\tableline
M$_{gas}$ & $1 \times 10^{10}$ M$_{\odot}$  \\
a$_{gas}$ & 7 kpc  \\
b$_{gas}$ & 0.4 kpc \\
\tableline
\end{tabular}
\end{center}
\end{table}

\begin{table*}
\begin{center}
\caption{Runs summary\label{tbl-runs}}
\begin{tabular}{c | c | c | c | c | c}
\tableline
Run & v$_{\rm ICM}$ (km/s) & P$_{\rm ICM}$ (dyne/cm$^{2}$) & T$_{\rm ICM}$ (K)& P$_{\rm ram}$ (dyne/cm$^{2}$) & t$_{\rm proj}$ (Myr)\\
\tableline
T80vh & 1900 & 4.2 $\times$ 10$^{-11}$ & 8.3 $\times$ 10$^{7}$& 11.6 $\times$ 10$^{-11}$ & 85\\
T3vh & 1900 &  4.2 $\times$ 10$^{-11}$ & 8.3 $\times$ 10$^{7}$& 11.6 $\times$ 10$^{-11}$ & 75\\
T3vl & 1413 & 2.66 $\times$ 10$^{-11}$ & 7.3 $\times$ 10$^{7}$& 5.29 $\times$ 10$^{-11}$ & 110\\
Sun et al. 2010 & & 1.8 $\times$ 10$^{-11}$ & $\sim$7 $\times$ 10$^{7}$& &  \\
\tableline
\end{tabular}
\end{center}
\end{table*}

Our galaxy model is the same as in TB10 and TB09, which used the spiral galaxy model described in Roediger \& Br\"uggen (2006).  We list the model parameters for our galaxy in Tables~\ref{tbl-const4} and \ref{tbl-gconst4}.  The stellar and dark matter components of the galaxy are static potentials.  As in TB09 and TB10, our galaxy position and box size allows us to follow gas 200 kpc above (in the wind, or z, direction) the galaxy.  To identify gas that has been stripped from the galaxy, we also follow a passive tracer which is initially set to 1.0 inside the galaxy (defined as gas that is above the ICM density) and $10^{-10}$ outside.  In the following analysis, we will use a minimum tracer fraction of 0.25 to find gas stripped from the galaxy (as in TB10).

\subsection{Introduction to the Three Runs}\label{sec:ICM}

The galaxy initially evolves in a static, high-pressure ICM with $\rho=$ 8.7 $\times$ 10$^{-28}$ g cm$^{-3}$ and $T = $ 9.069 $\times$ 10$^6$ K ($\rho=$7.57 $\times$ 10$^{-28}$ g cm$^{-3}$ and $T = $ 1.04294 $\times$ 10$^7$ K for the slower wind case), to allow cool, dense gas to form in the galaxy (each of our three runs has about 3 $\times$ 10$^9$ M$_\odot$ of gas with densities at or above 10$^{-22}$ g cm$^{-3}$ when the wind hits the disk).  This naturally generates a multiphase ISM (see Tasker \& Bryan 2006 and TB09 for more discussion of the ISM properties).  

After 155 Myrs, we reset the boundary conditions to generate a constant ICM inflow along the inner z-axis, which is always face-on to the galaxy.  In Table \ref{tbl-runs} we show the details of each of the three runs.  The table includes the ICM parameters and the time after the wind has hit the galaxy at which the X-ray tail is 80 kpc long (this is how we choose the outputs to compare to the observations of ESO 137-001).  `T80' indicates cooling to 8,000 K and `T3' indicates cooling to 300 K, while `vh' and `vl' indicate high and low velocity wind, respectively.  See TB09 for other details regarding the general numerical setup, and TB10 for a discussion of the general impact of the cooling floor on the tail structure.  In order to compare with both observations and our previous work, we use a face-on wind direction.

\subsection{Projections}\label{sec:projection}

\textit{Enzo} outputs the density and temperature of the gas in each cell.  To transform these values into \ion{H}{I} column density and H$\alpha$ intensity, we used Cloudy, version 08.00 of the code last described by Ferland et al. (1998).  Using a table of temperatures and densities, we calculated the hydrogen neutral fraction and H$\alpha$ emissivity.  This table is then used to calculate the observational quantities for each cell, which are then summed to generate an image.  This is described in detail in TB10, but briefly, we included CMB radiation, the cosmic ray background, bremsstrahlung radiation from the ICM and the 2005 version of the Haardt \& Madau (2001) $z=0$ metagalactic continuum, as implemented by Cloudy.  

We chose to calculate the neutral fraction and H$\alpha$ emissivity for a plane-parallel gas cloud of width 100 pc.  We selected this width because it loosely corresponds to the cell size of most of the gas in the tails, and accounts approximately for attenuation of the ionizing background radiation.  If we assumed the radiative thin limit (by using a very small cloud size in Cloudy), it would somewhat decrease the amount of \ion{H}{I} we predict, and increase the H$\alpha$ emission for dense, low-temperature gas.  We discuss the use of different cloud sizes in detail in TB10, however, in the simulations in this paper we find that using a 10 pc cloud does not significantly change our H$\alpha$ flux nor our \ion{H}{I} column densities because of the higher densities in our tail gas.  This indicates that radiative transfer effects are not that important for this work.

To create X-ray surface brightness projections, we use a spectral lookup table that depends on temperature and density, assuming a constant metallicity of 0.3 solar, as computed using a Raymond-Smith code (Raymond \& Smith 1977), as updated in XSPEC (Arnaud 1996).  The X-ray band we use is 0.5 keV to 2.0 keV, following Sun et al. (2006).  

\section{Comparison to Observations}

In this section, we compare our simulated stripped tail to observations of ESO 137-001.  We choose an output time at which the X-ray tail length is about 80 kpc in order to match the observations of Sun et al. (2010); this time is shown in Table \ref{tbl-runs}.  We are using this comparison with the observations of ESO 137-001 to better understand the physics at work in producing X-ray, H$\alpha$, and \ion{H}{I} tails, so we have not tuned our simulation specifically to find an exact match to the galaxy and tail of ESO 137-001.  While our ICM parameters are similar to those of the ICM near ESO 137-001, they are not the same.  The final row in Table \ref{tbl-runs} displays the ICM parameters from the observations of Sun et al. (2010).  We have not attempted to model the exact angle between the galaxy's disk and orbital motion (Woudt et al. (2008) find a position angle of 125$^\circ$), and instead use a face-on wind, and we are using a larger galaxy (Initially our galaxy has a radius of 26 kpc, and in the comparison projections the radius is about 15 kpc, while the 2MASS isophotal radius for ESO 137-001 is 6.1 kpc (Skrutskie et al. 2006)).  In the following comparisons we will take note of the differences between our simulations and the observations of Sun et al. (2006; 2007; 2010).

\begin{figure*}
\includegraphics[scale=1.55,trim= 7mm 6mm 15mm 50mm,clip]{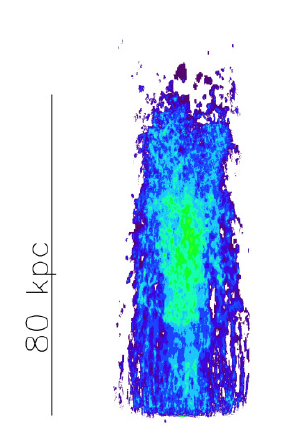}
\includegraphics[scale=1.55,trim= 7mm 6mm 15mm 50mm,clip]{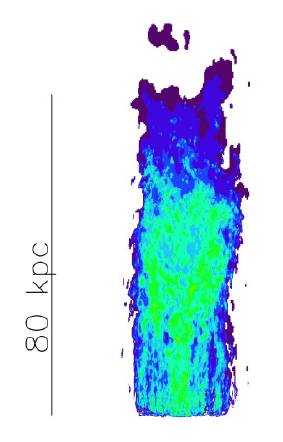}
\includegraphics[scale=1.55,,trim= 7mm 6mm 15mm 50mm,clip]{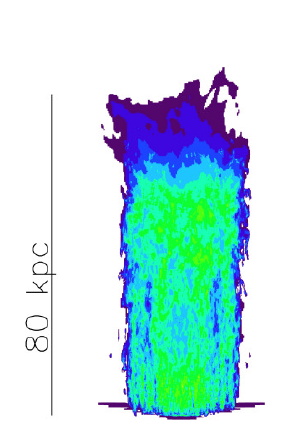}
\includegraphics[scale=0.55]{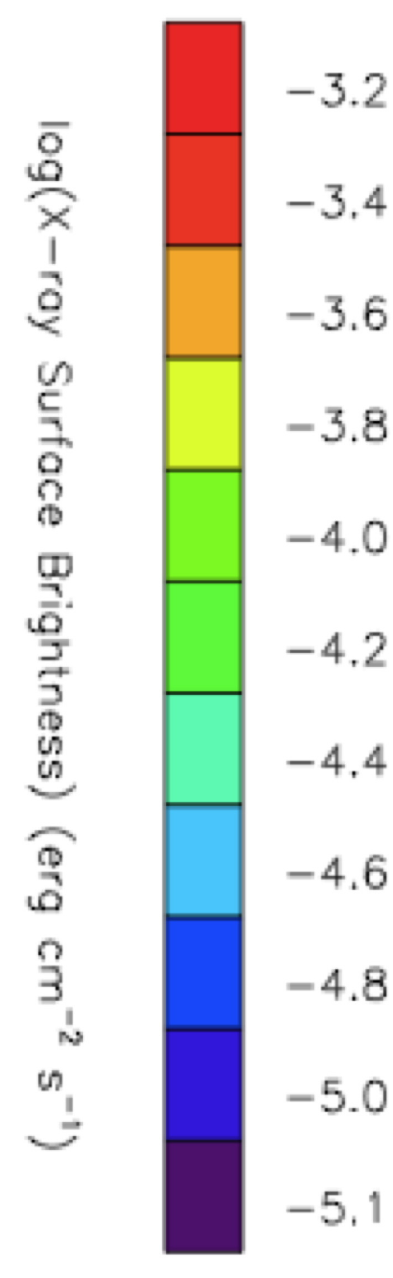}
\caption{X-ray surface brightness maps of our three simulated galaxies, seen side-on.  From left to right, we show T3vl, T80vh, T3vh.  We have chosen the outputs with tail lengths of 80 kpc for our comparison to the observations of Sun et al. (2006; 2007; 2010).  See discussion in Section \ref{xraycomp}}
\label{fig:xrayproj}
\end{figure*}

We have chosen the output time at which each simulation has an X-ray tail whose length is about 80 kpc.  This means that the length of time that each galaxy has been stripped in the different simulations is not the same.  In Table \ref{tbl-tempmass}, we list the amount of gas in the tail in three different temperature ranges.  These masses include all of the gas above 10 kpc from the disk with tracer fractions of at least 25\%,  so some of the gas included is at low density, as shown in Figure \ref{fig:rhot} (too low to be observable).  Nevertheless, it can be very loosely stated that longer stripping times may result in more mixing of hot gas into the ICM, and in more dense gas being stripped and condensing into clouds in the tail.  This can be seen by comparing T80vh or T3vl with T3vh (of course there are also other differences in the simulations that influence the tail gas, such as cooling floor and ICM pressure).  However, we stress that in this paper we mainly focus on what causes a tail to be bright in X-ray and H$\alpha$ emission, so focus more on ICM pressure than the amount of time a galaxy has been stripped (although we do consider this timescale in our discussion of heat conduction).

\begin{table}
\begin{center}
\caption{Amount of Stripped gas at Different Temperatures\label{tbl-tempmass}}
\begin{tabular}{c | c | c | c}
\tableline
Run & T $<$ 10$^4$ K & 10$^4$ $<$ T $<$ 10$^5$ K & 7 $\times$ 10$^5$ $<$ T $<$ 4 $\times $10$^7$ K\\
 & 10$^9$ M$_\odot$ & 10$^9$ M$_\odot$ & 10$^9$ M$_\odot$ \\
\tableline
T80vh & 1.4 & 1.6 & 6.0 \\
T3vh & 1.1 & 1.2 & 9.1 \\
T3vl &  4.2 & 2.9 & 5.9 \\
\tableline
\end{tabular}
\end{center}
\end{table}

\subsection{X-ray}

\subsubsection{Comparison to ESO 137-001}\label{xraycomp}

We first compare the X-ray characteristics of our simulations to observations.  The X-ray surface brightness projections of our runs are shown in Figure \ref{fig:xrayproj}.  As in Sun et al. (2006; 2010), we measure the X-ray emission between 0.5-2.0 keV.  The surface brightness profiles along the tails are in rough agreement with that of ESO 137-001, which is bright near the disk and has a second bright region about 40 kpc from the disk before becoming less luminous to the end of the tail.  Our model does not reproduce the exact X-ray surface brightness profiles:  we do not have a model that includes both the bright emission near the disk and a surface brightness decrease across the entire tail before the second brightness peak. 

\begin{table*}
\begin{center}
\caption{X-ray tail attributes\label{tbl-xtail}}
\begin{tabular}{c | c | c | c | c}
\tableline
Run or  & \textit{l $\times$ w} & L$_{0.5-2 \rm{keV}}$ & L$_{0.5-2}$ {\rm corrected}$^{\tablenotemark{a}}$ & T  \\
Observation & (kpc $\times$ kpc)& (10$^{40}$ erg s$^{-1}$) & (10$^{40}$ erg s$^{-1}$) & (10$^7$ K)\\
\tableline
T80vh & 80 $\times$ 26 & 66.3 & 4.8 & 2.5\\
T3vh & 80 $\times$ 30 &  93.0 & 5.1 & 2.1 \\
T3vl & 80 $\times$ 30 & 41.9 & 3.6 & 1.4 \\
Sun et al. 2010 & 80 $\times$ 8, 80 $\times$ 7 & (8.3 $\pm$ 0.4) & (8.3 $\pm$ 0.4) & 0.93 $\pm$ 0.05 \\
\tableline
\end{tabular}
\tablenotetext{a}{The corrected luminosity multiplies the simulated X-ray luminosity by the ratio between the simulated and observed tail volumes and the ratio between the simulated and observed ICM thermal pressures (see text).}
\end{center}
\end{table*}

To make the X-ray projections, we adopted a minimum observable surface brightness of 7.1 $\times$ 10$^{-6}$ erg cm$^{-2}$ s$^{-1}$, which we estimated from the total luminosity of the observed tail and brightness profiles in Sun et al. (2010) (our cluster background is about 3.6 $\times$ 10$^{-6}$ erg cm$^{-2}$ s$^{-1}$).  This results in a luminosity difference of an order of magnitude between the lowest and highest surface brightness features in all of our simulated tails, similar to that in the observed tail of ESO 137-001. 

Our tails are narrow and show a nearly constant width along the entire tail, in agreement with the X-ray morphology reported in Sun et al. (2006; 2010).  This is in contrast to simulations that do not include radiative cooling and produce flared tails (e.g. Roediger, Br\"uggen \& Hoeft 2006)  However, it is clear that we do not have separated X-ray tails as in the observations.  This is not surprising given the explanation of Sun et al. (2010) that the two tails likely result from the stripping of two spiral arms.  As we discuss in detail in TB09, our disks fragment but do not form spiral arms. 

In Table \ref{tbl-xtail}, we list some of the characteristics of our simulated X-ray tails to compare with the observed tail.  To calculate the luminosity of the tail, we find the total energy emitted and subtract the background emission from the ICM.  As the table demonstrates, the simulated tails are wider, more luminous, and have a higher average (emission-weighted) temperature than the (spectroscopic) temperature measured by Sun et al. (2010).   We now address these differences.

The tail is most likely wider because we are modeling a large spiral galaxy, while ESO 137-001 is thought to be a $\sim$ 0.2L$_*$ galaxy, with a smaller galactic radius (Sun et al. 2006 and references therein).  In fact, the entire volume of the tail in any of our three runs is nearly an order of magnitude larger than the tail of ESO 137-001 (we assume that the tails are cylindrical with the heights and diameters denoted in Table \ref{tbl-xtail}).  The other difference is that our ICM pressure is also somewhat larger than calculated in Sun et al. (2010) by either a factor of 1.47 (T3vl) or 2.33 (T3vh and T80vh).  

These differences in tail volume and ICM pressure impact the observables in two ways.  First, as we discuss in TB10 and below, the compression of stripped gas by the ICM determines the temperature and density distribution of gas in the tail.  Therefore, a higher ICM pressure results in higher-density hot gas ($T > 10^6$ K), which will have a higher X-ray emissivity.  As we argue in the next section, this makes the X-ray luminosity proportional to the ICM pressure.  Second, if we assume that the filling factor of X-ray emitting gas is the same in our simulations as in the tail of ESO 137-001, then the total luminosity is also directly proportional to the volume of the entire tail.  The X-ray luminosity after applying these corrections is also shown in Table \ref{tbl-xtail} (assuming X-ray luminosity is directly proportional to both the ICM pressure and the volume of the tail).  When we account for these differences, our measured luminosities are within about a factor of two of the X-ray luminosity of ESO 137-001.  

As shown in Table~\ref{tbl-xtail}, there is also a factor of two difference between the temperatures of the simulated and observed X-ray tails; however, this is most likely due to the different ways in which this quantity is determined in simulations as compared to observations.  We use luminosity-weighted temperatures, which tend to be higher than spectroscopic temperatures when there is a range of gas temperatures (see Mazzotta et al. 2004). Sun et al. (2010) discuss how the spectral fitting of the iron-L hump biases their temperatures low if there are significant emission components at \textit{k}T = 0.4 - 2 keV, which is certainly the case in our simulation and in the tail of ESO 137-001 (Sun et al. 2010; Sivanandam et al. 2009).  Therefore, the agreement is probably considerably better than indicated in Table~\ref{tbl-xtail}.

\begin{figure*}
\begin{center}
\includegraphics[scale=0.8,trim= 1mm 90mm 10mm 75mm,clip]{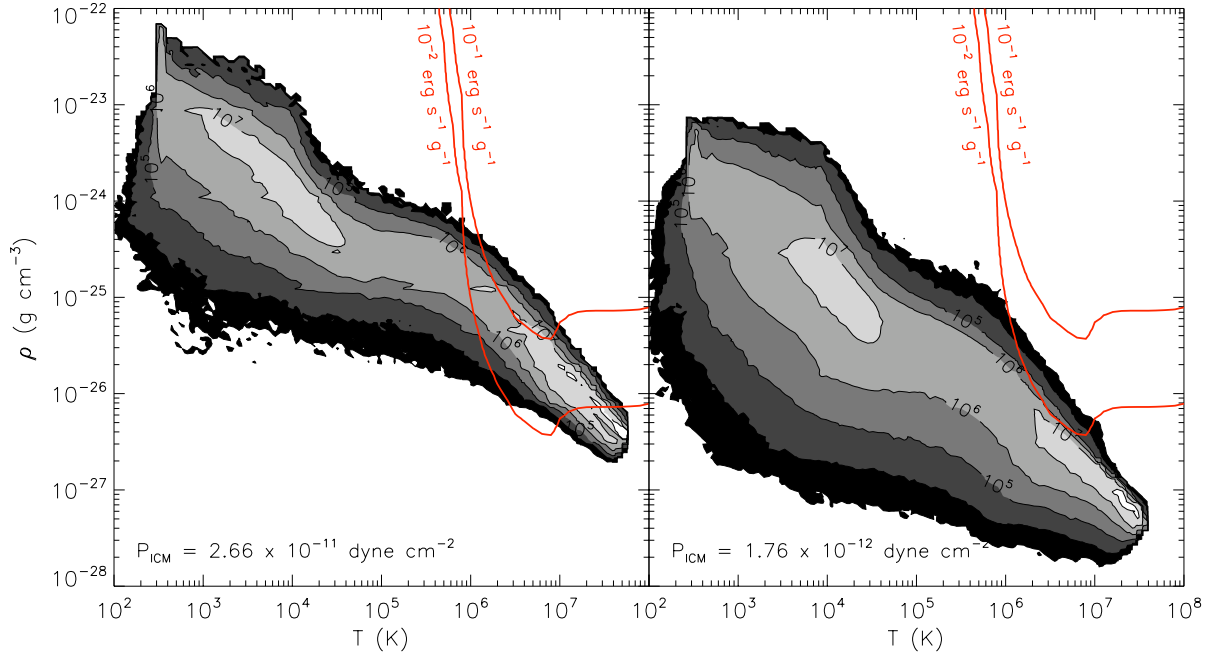}
\end{center}
\caption{Contour plots showing the mass in gas at different densities and temperatures for the T3vl run on the left, $\tmin = 300$ K (from TB10) on the right.  The contours are spaced by a factor of 10 in mass.  For a gas cell to be included in the contour plot, it must have at least 25\% of its mass originating from the galaxy (based on the tracer fluid).  The two curves denote lines of constant X-ray luminosity per mass (or emissivity per density).  This figure illustrates that high ICM pressure produces an X-ray bright tail.  See discussion in Section \ref{xrayexplain}}
\label{fig:rhot}
\end{figure*}

\subsubsection{What Makes a Tail X-ray Bright?}\label{xrayexplain}

In this section, we examine what lights up an X-ray bright tail, and explain why some observed tails, like ESO 137-001, exhibit X-ray emission, while others do not.  In Figure \ref{fig:rhot} we show the mass-weighted distribution of density and temperature of gas in the wake that originated from the galaxy from our T3vl run (left) and from the $\tmin = 300$ K run from TB10 using the corrected cooling curve (right).  The plots include all of the gas located between 10 kpc and 240 kpc above the disk that has at least 25\% of it's mass originating in the galaxy (as determined by the tracer fraction).  We choose to highlight these two cases because the ICM pressure is the only difference between the two simulations (the thermal pressures differ by a factor of 15).  As discussed in more detail in TB10, the ($T > 10^5$ K) gas in the wake is largely in pressure equilibrium and so falls roughly along a line of constant pressure.  The impact of the higher ICM pressure can be seen as a shift to higher density and temperature in the left panel.

\begin{figure*}
\begin{center}
\includegraphics[scale=10.5,trim= 21mm 8.75mm 23.1mm 84.9mm,clip]{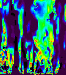}
\includegraphics[scale=10.5,trim= 21mm 8.75mm 23.1mm 84.5mm,clip]{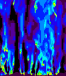}\\
\includegraphics[scale=10.5,trim= 21mm 8.75mm 23.1mm 84.5mm,clip]{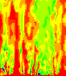}
\includegraphics[scale=10.5,trim= 21mm 8.75mm 23.1mm 84.5mm,clip]{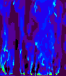}
\end{center}
\caption{Projections of a very thin (0.3 kpc) slice from the T80vh run (images are 16 kpc $\times$ 19 kpc).  The upper left panel shows X-ray surface brightness (red brightest), which demonstrates that X-ray emitting gas is associated with diffuse, mixed gas, but not necessarily material recently stripped from dense clouds in the tail.  The lower left panel (temperature) demonstrates that the brightest X-ray emitting gas is associated with intermediate temperatures.  Similarly, the upper right panel shows that H$\alpha$ emission is produced primarily at the edges of dense clouds.  The lower right panel (gas surface density) demonstrates that the brightest H$\alpha$ is produced in high surface density clouds.}
\label{fig:slices}
\end{figure*}

In red we also plot two contours of constant luminosity per mass (or emissivity per density).  The lower contour is at 10$^{-2}$ erg s$^{-1}$ g$^{-1}$.  If the minimum observable X-ray surface brightness is 10$^{-5}$ erg s$^{-1}$ cm$^{-2}$, then in order for X-ray emission to reach this level, there must be a sufficient amount of gas at or above the red contour.   More precisely, we need a column density of hot gas of 10$^{21}$ cm$^{-2}$ in order to produce observable emission (this corresponds to about 10$^4$ M$_\odot$ along a single line of sight through our 38 pc $\times$ 38 pc cells).  In general, only the colder gas (T $\le$ 10$^4$ K) in the tail has these high surface densities (10$^{21}$ cm$^{-2}$), so we also plot an emissivity line which only requires a surface density of 10$^{20}$ cm$^{-2}$ in order to be observable (the upper 10$^{-1}$ erg s$^{-1}$ g$^{-1}$ contour).  Gas that is hot enough to emit X-rays (hotter than $\sim$7 $\times$ 10$^5$ K) has this lower surface density in the tail (10$^{20}$ cm$^{-2}$).

This figure shows that gas at higher densities and lower temperatures than the ICM (but above $\sim$ 10$^6$ K) will be emitting the most strongly in X-rays.  As galactic gas is stripped it either cools into clouds ($T <  10^5$ K) or is compressed to the ICM pressure and begins to mix with ICM gas.  While the tail gas is at the high ICM pressure -- but before it is completely mixed with the ICM -- it will be X-ray bright.  The mixing occurs along the line of constant pressure seen in Figure~\ref{fig:rhot}.  

The separation between cold clouds and X-ray emitting gas is seen clearly in the left panel of Figure \ref{fig:slices}, which shows a thin slice from a detailed section above the disk.  The cold clouds are black in this figure, indicating no X-ray emission, while the X-ray emitting gas is not confined to regions close to the dense clouds.  Most of the X-ray bright cells in our simulation have between 70\% and 90\% of their gas originating from the galaxy.  In the lower left panel of Figure \ref{fig:slices} we show the gas temperature, which when compared with the X-ray slice again highlights that while some mixing and radiative cooling can enhance the X-ray emission, but too much lowers the density of the hot gas to the point where the X-ray emissivity is low.  The X-ray bright gas may have either been stripped as hot gas from the disk and slowly mixed, or stripped from the dense clouds in the tail and mixed into the ICM.    

\begin{figure*}
\includegraphics[scale=1.55,trim= 7mm 6mm 15mm 50mm,clip]{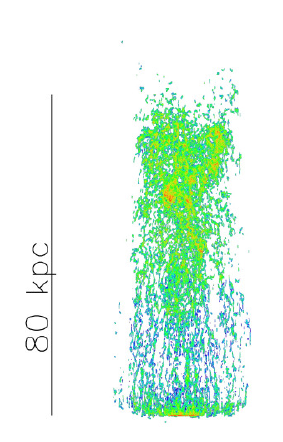}
\includegraphics[scale=1.55,trim= 7mm 6mm 15mm 50mm,clip]{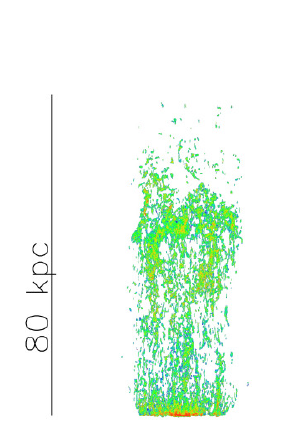}
\includegraphics[scale=1.55,,trim= 7mm 6mm 15mm 50mm,clip]{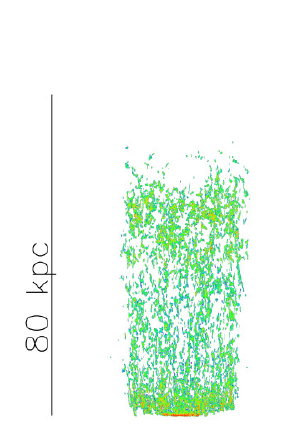}
\includegraphics[scale=0.55]{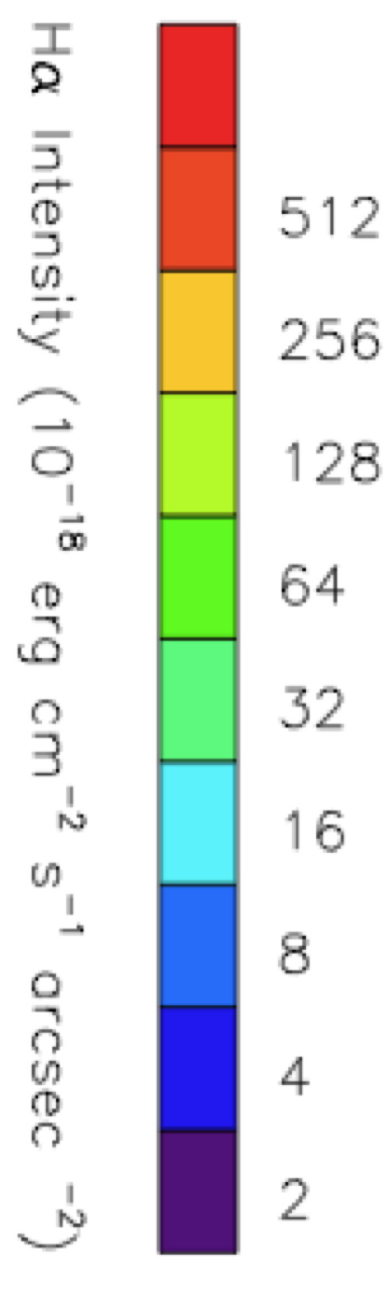}
\caption{H$\alpha$ intensity projections.  These can be compared with the diffuse emission observed by Sun et al. (2007).  From left to right: T3vl, T80vh, T3vh.  See Section \ref{sec:ha} for discussion.}
\label{fig:haproj}
\end{figure*}

\subsection{H$\alpha$}\label{sec:ha}

Next, we turn to H$\alpha$ emission, shown in Figure~\ref{fig:haproj}.  The minimum observable surface brightness that we adopt for these maps is 2 $\times$ 10$^{-18}$ erg s$^{-1}$ cm$^{-2}$ arcsec$^{-2}$ (see Sun et al. (2007) and references therein).  As described earlier, we use Cloudy to determine the H$\alpha$ emissivity given a gas temperature and density.  

We find, as shown in Figure \ref{fig:haproj}, highly structured, long tails of H$\alpha$ emission.  Note that we do not include UV radiation from star formation or AGN (except from the metagalactic background, as described in section \ref{sec:projection}).  Figure \ref{fig:slices} shows that H$\alpha$ emission peaks around the edges of cold clouds, which can be seen by comparing the upper right panel of H$\alpha$ emission with the lower right panel showing gas surface density.  

\begin{figure*}
\includegraphics[scale=1.55,trim= 7mm 6mm 15mm 50mm,clip]{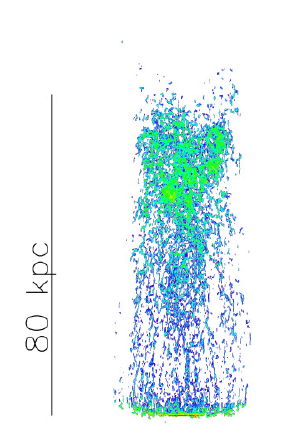}
\includegraphics[scale=1.55,trim= 7mm 6mm 15mm 50mm,clip]{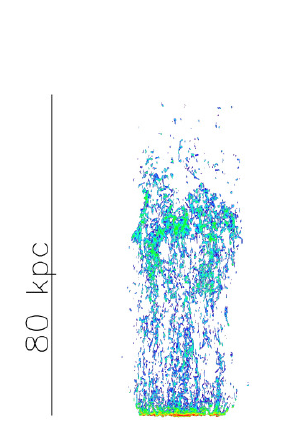}
\includegraphics[scale=1.55,,trim= 7mm 6mm 15mm 50mm,clip]{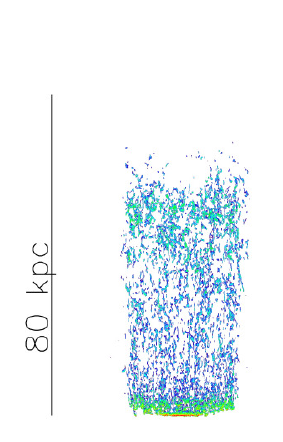}
\includegraphics[scale=0.55]{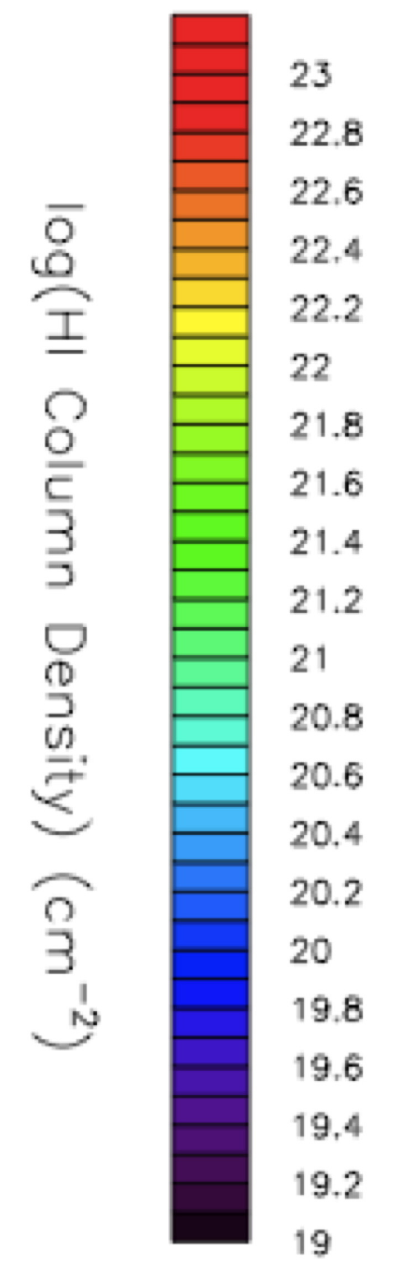}
\caption{\ion{H}{I} column density projections at the maximum resolution of our simulation (38 pc).  From left to right: T3vl, T80vh, T3vh.  See Section \ref{sec:h1} for discussion.}
\label{fig:hiproj}
\end{figure*}

The $\tmin=8000$ K and $\tmin = 300$ K runs show a similar amount of H$\alpha$ emission.  This is because, while the minimum temperature affects the temperature in the central regions of the clouds, it does not strongly affect the cloud edges in the simulation, whose characteristics are more determined by the interaction between the cloud and the ICM.  We find that changing the cloud size parameter in our Cloudy run has only a small effect on the total H$\alpha$ flux (less than 10\% in all three runs).  This is because most of our emission is produced by collisional processes rather than photoionization, and so is mostly dependent on the gas temperature, not the optical depth to ionizing photons (recall that we do include star formation in the simulation and so do not model HII regions within the tail).

\begin{table}
\begin{center}
\caption{H$\alpha$ tail attributes\label{tbl-hatail}}
\begin{tabular}{c | c | c | c}
\tableline
Run or & \textit{l $\times$ w} & f$_{\rm{H}\alpha}/10^{-14}$ & f$_{\rm{H}\alpha}/10^{-14}$ {\rm corrected}$^{\tablenotemark{a}}$ \\
Observation & (kpc $\times$ kpc) & (erg s$^{-1}$ cm$^{-2}$) & (erg s$^{-1}$ cm$^{-2}$)\\
\tableline
T80vh & 67 $\times$ 27 & 75.8 & 2.23\\
T3vh &  65 $\times$ 29 & 75.9 & 2.00 \\
T3vl &  80 $\times$  30 & 97.9 & 1.96 \\
Sun et al. (2007) & $\sim$40 $\times$ 6 & 4.4 & 4.4\\ 
\tableline
\end{tabular}
\tablenotetext{a}{The corrected flux multiplies the simulated H$\alpha$ flux by the ratio between the simulated and observed tail volumes.}
\end{center}
\end{table}

Because we do not include star formation, we cannot compare our H$\alpha$ emission to the 33 \ion{H}{II} regions seen by Sun et al. (2007).  We do compare the total flux from our tail to the observed diffuse H$\alpha$ emission in Table \ref{tbl-hatail}.  Again, the volume of our diffuse tail is much larger than that observed by Sun et al. (2007), and we show the corrected H$\alpha$ flux by dividing by the volume ratio in the table.  The difference in tail widths is from the difference in galaxy sizes, while the length of the observed tail is the minimum length of the diffuse emission because there is a bright star in the field (Sun et al. 2007).  When we take the different volumes of the tail into account, our simulated H$\alpha$ flux is less than that observed from the tail of ESO 137-001, but in all three cases, the simulated H$\alpha$ flux is within a factor of 2.5 of the observed flux. 

There are a number of possible explanations for our slightly lower predicted H$\alpha$ flux, including the possibility that some unresolved \ion{H}{II} regions are counted as diffuse flux in the observations.  Other heating sources such as thermal conduction --- an effect we do not include in the simulations --- could give rise to more H$\alpha$ flux than we see in our simulations.  We also find that numerical resolution plays a role in the total H$\alpha$ emission, as we will discuss in Section~\ref{sec-resolution}. 

\begin{figure}
\begin{center}
\includegraphics[scale=1.55,trim= 7mm 6mm 15mm 50mm,clip]{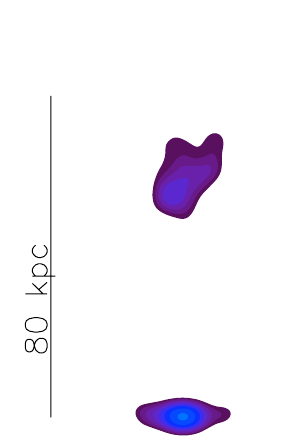}
\includegraphics[scale=0.55,trim= 0mm 5mm 0mm 0mm,clip]{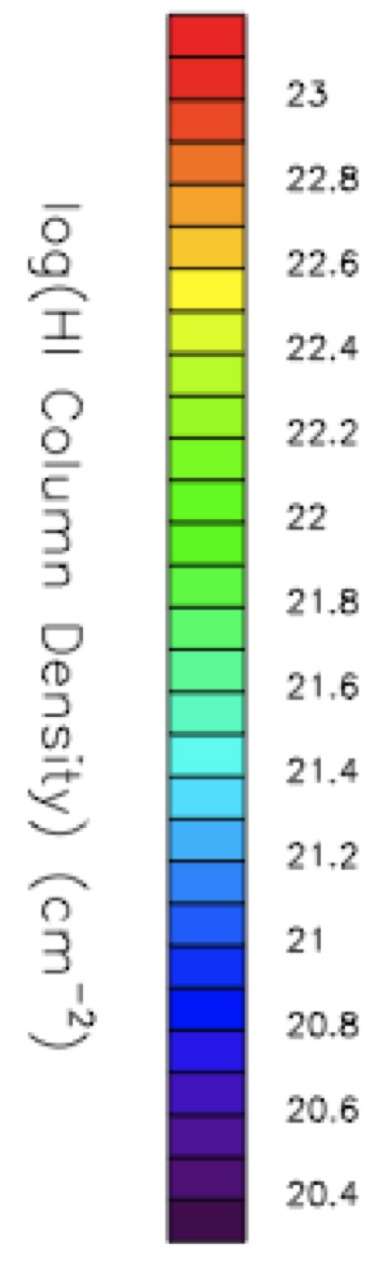}
\end{center}
\caption{\ion{H}{I} column density projection smoothed to the resolution of the observations by Vollmer et al. (2001) using ATCA.  The only run with observable \ion{H}{I} in the tail using 30" resolution and a minimum column density of 2 $\times$ 10$^{20}$ is T3vl.  See Section \ref{sec:h1} for discussion.}
\label{fig:ATCAV01}
\end{figure}

\subsection{\ion{H}{I}}\label{sec:h1}

In this section we consider \ion{H}{I} column density, comparing projections of our simulations to observations.  In Figure \ref{fig:hiproj} we show the \ion{H}{I} column density at the best resolution of our simulation.  As discussed in Section \ref{sec:projection}, we use Cloudy to determine the neutral fraction given a temperature and density (with the assumed metagalactic UV flux) and then apply that value in our projection routine.  

In Figure \ref{fig:ATCAV01} we only show a projection comparable to the observations performed by Vollmer et al. (2001), using a resolution of 30" and a minimum column density of 2 $\times$ 10$^{20}$ cm$^{-2}$.  We also correct for the difference in disk sizes by dividing our column densities by a factor of 3 (the ratio of X-ray tail diameters).  Only the tail of T3vl would have been observed by Vollmer et al. (2001).  This is because in the slower velocity case there is a smaller velocity difference between the clouds and the ICM wind and the ICM is slighty less dense (83\% of the ICM density in the higher velocity cases), which results in less cloud ablation and more high density gas (both a slightly higher maximum density of gas and slightly more gas at any particular density above $n \sim 1$ cm$^{-3}$).  This difference between the non-detection by Vollmer et al. (2001c) and our prediction that T3vl would be detected implies that ESO 137-001 is most likely moving more quickly through the ICM than 1413 km s$^{-1}$.    The \ion{H}{I} column densities in T3vh and T80vh are within a factor of 3 and 1.3 respectively below the maximum column density as determined by the non-detection of Vollmer et al. (2001).  

Figure \ref{fig:slices} clearly shows that bright H$\alpha$ emission is produced only at the edges of dense neutral clouds.  This spatial correspondence between neutral and H$\alpha$ emitting gas agrees with the observations of Sivanandam et al. (2009), who found molecular hydrogen in the tail to the farthest distance they could search--20 kpc.  We predict that with a deeper observation \ion{H}{I} will be observed to at least 40 kpc (the length of the observed H$\alpha$ tail), and likely even farther (our shortest \ion{H}{I} tail is 65 kpc in T3vh), unless heat conduction is quite efficient.

\section{Discussion}

\subsection{Which Stripped Galaxies will have X-ray Tails?}

In this section, we comment on the likelihood of finding more X-ray tails.  In order for a tail to have observable X-ray emission, it must be in a high pressure ICM.  We can estimate the pressure necessary to produce bright X-ray emission simply by shifting our mass-weighted distribution of gas until the lowest mass contour lies along the upper 10$^{-1}$ erg s$^{-1}$ g$^{-1}$ contour.  We choose this contour because in every simulation we have run, projections of surface density have shown that most of the tail has a column density of at least 10$^{20}$ cm$^{-2}$.  Gas in the tail with higher surface densities tends to be in cold clouds (T $<$ 10$^4$ K) (see TB10), while X-rays are emitted by hot (T $>$ 10$^6$ K), diffuse gas with lower column densities (Figure \ref{fig:rhot}).  This method results in a minimum ICM pressure of 9 $\times$ 10$^{-12}$ erg cm$^{-3}$.  

In Table \ref{tbl-clusterrad} we show the cluster radius at which the ICM pressure falls below our minimum value.  In general, we calculated the cluster radius using a $\beta$-model and constant temperature from the papers cited in column 3.  There are three exceptions to this.  First, the X-ray intensity contours in Figure 17 of Wang et al. (2004) denote the highest density ICM region (and are not spherically symmetric), which is the only region of A2125 with a high enough density to produce our minimum ICM pressure for X-ray bright tails.  Second, we find the radius of Perseus with our minimum ICM pressure directly from Figure 9 in Ettori et al. (1998).  Finally, we use the entropy measurements of Cavagnolo et al. (2009) to find that as close as 100 kpc from M87 the ICM pressure is below 9 $\times$ 10$^{-12}$ erg cm$^{-3}$.  Note that these are first estimates generally using spherically symmetric cluster profiles, which oversimplify the structure of the ICM.
We can now comment on whether observed X-ray tails are in high pressure ICMs (defined here as at or above 9 $\times$ 10$^{-12}$ dyn cm$^{-2}$).  The two X-ray tails in A3627, ESO 137-001 and ESO 137-002, are both within the high pressure ICM (Sun et al. 2007; Sun et al. 2010).  However, Sun et al. (2010) find that using their spectroscopic temperatures, the tail gas is over-pressured with respect to the ICM.  This might indicate that the spectroscopic temperature does not correctly model the tail temperature.  We find that turbulent pressure is not strong enough to greatly increase the gas pressure in the tail (Figure \ref{fig:rhot} and TB10).  

C153 is near the X-ray peak of A2125 (Wang et al. 2004).  UGC 6697 in A1367 is at least 450 kpc from the X-ray peak of the cluster, outside of our calculated high-pressure radius.  However, it is near an infalling subcluster and in a higher-density ICM than predicted using the $\beta$-model centered on the X-ray peak of A1367.  Using the fit to the subcluster by Donnelly et al. (1998), Sun \& Vikhlinin (2005) calculated the surrounding pressure to be 7.9 $\times$ 10$^{-12}$ dyn cm$^{-2}$, very close to our minimum pressure for observable X-ray emission.

\subsection{Efficiency of Heat Conduction}

\begin{table}
\begin{center}
\caption{Where in clusters will tails be X-ray bright?\label{tbl-clusterrad}}
\begin{tabular}{c | c | c }
\tableline
Cluster & Outer radius (kpc) & Reference\\
\tableline
A3627 & 250 \textit{h$_{50}^{-1}$} & Bohringer et al. (1996); \\
 & & Sun et al. (2010)\\
A2125 & 28.5 - 89 \textit{h$_{71}^{-1}$} & Wang et al. (2004)\\
A1367 & 274 \textit{h$_{50}^{-1}$} & Mohr et al. (1999) \\
CL 0024+0016 & 250 \textit{h$_{70}^{-1}$} & Zhang et al. (2005)\\
Coma & 613 \textit{h$_{50}^{-1}$} & Briel et al. (1992) \\
Perseus & 2000 \textit{h$_{50}^{-1}$} & Ettori et al. (1998) \\
Virgo (M87) & $<$ 100 \textit{h$_{70}^{-1}$} & Cavagnolo et al. (2009)\\
\tableline
\end{tabular}
\end{center}
\end{table}

Heat conduction, which we do not include in these simulations, could be important for the survival of cool clouds in the ICM and H$\alpha$ emission; however, it can be suppressed by magnetic fields (e.g. Vollmer et al. 2001).  If it is an efficient way to transport heat from the ICM to cold, stripped gas, then the survival time of \ion{H}{I} clouds would be less than predicted in this paper, and the length of the tails would be shorter.  We can estimate the efficiency of heat conduction by comparing an analytic calculation of the evaporation time of the most dense clouds in our simulations to the length of time we expect clouds have survived in order to produce the observations of ESO 137-001.

First, we estimate the evaporation time for a typical cloud if heat conduction is not suppressed, and we will define this calculated evaporation time as the time for cloud evaporation if heat conduction is 100\% efficient.  We follow Cowie \& McKee (1977), as in Vollmer et al. (2001).  In our clouds, we find that the mean free path for ions is comparable to or greater than the temperature scale length, so we need to use the saturated heat flux equations.  

Evaporation time is proportional to $f^{-1} r_{\rm cloud}^{11/8} n_{\rm cloud}$ $T_{\rm ICM}^{-5/4} n_{\rm ICM}^{-11/8}$, where $f$ is the conduction efficiency relative to Spitzer.  Solving for the evaporation time for each of our three runs, we find that using a cloud radius of 100 pc, and the maximum gas density in our tails at the time of our comparison, the evaporation times of our three runs are $\sim 4$ Myr for T80vh, $\sim 8$ Myr for T3vh, and $\sim 10$ Myr for T3vl.  We choose to use the highest density found in our tails because we want to calculate the longest plausible evaporation time in order to find the most conservative estimates for the maximum efficiency of heat conduction.  The $\tmin=300$ K runs have a maximum density in their tails of 2 $\times$ 10$^{-23}$ g cm$^{-3}$ and T80vh has a maximum density of 10$^{-23}$ g cm$^{-3}$.

We have shown that diffuse H$\alpha$ emission directly traces neutral clouds (Figure \ref{fig:slices}), which means that the dense clouds must not be entirely evaporated before they reach 40 kpc above the disk.  We are also able to measure the velocity of gas in our tails (see TB10 for details), so can estimate the time it would take for a cloud to reach 40 kpc above the disk.  We use generous estimates of 1200 km s$^{-1}$ for the high velocity cases and 900 km s$^{-1}$ for the lower velocity run in order to calculate the shortest time it would take these clouds to reach 40 kpc above the disk (which will maximize the value we find for the efficiency of heat conduction).  Therefore, the efficiency of heat conduction must be less than 24\% (T3vl and T3vh), or 12\%(T80vh).  

Once there is a deep observation of the \ion{H}{I} tail, we will be able to compare the length of ESO 137-001's tail to the lengths of our simulated tails to find a {\it minimum} efficiency of heat conduction.  To do this we will use a similar argument to the one above; namely, compare the amount of time a cloud takes to reach its height above the disk to the evaporation time, assuming that heat conduction is responsible for destroying the clouds.  It is important to consider our simulations in this calculation, because they show that even without heat conduction, \ion{H}{I} may not extend along the entire length of the X-ray tail.  

The simulations also show the limitations of using a simple velocity and distance argument to calculate the length of time a galaxy has been stripped--using 900 km s$^{-1}$ and the length of the \ion{H}{I} tail we find that the galaxy in T3vl has been stripped for 87 Myr, while the actual time after the wind has hit the galaxy is 110 Myr.  The high velocity cases have an even larger discrepancy between the actual stripping time and the time calculated using the \ion{H}{I} tail length.  

Despite these caveats, we can make a very rough first estimate using Figure \ref{fig:ATCAV01} (T3vl) and the non-detection result of Vollmer et al. (2001).  The \ion{H}{I} in our smoothed projection begins about 50 kpc above the disk.  If the non-detection of Vollmer et al. (2001) means that there are no neutral clouds at this height above the disk, then we can calculate a minimum efficiency for heat conduction, which is 18.5\%.  However, as we discussed above, we expect that a deeper observation will find \ion{H}{I} beyond this height above the disk.  


\section{Numerical Considerations}\label{sec-numerical}

\subsection{Radiative Cooling Floor}\label{sec-cool}

As noted earlier, we used two different values for our radiative cooling floor (8000 K and 300 K) in order to explore in a simple way the potential impact of processes which we do not include in the simulation.  We discuss the physical motivations for the two cooling floors in TB10.  We find that the change in the minimum temperature makes only a relatively minor change in the morphology of the flow, and doesn't significantly change the resulting length of the tail.  This also translates into only a relatively small difference in the predicted observables, with changes of less than 30\% in both X-ray and H$\alpha$ luminosities.\footnote{Note that this differs from the conclusion reached in TB10, where we found that the $\tmin=300$ K run predicted lower H$\alpha$ emission by more than an order of magnitude.  This incorrect conclusion was reached because of the cooling curve error for that run discussed in section~\ref{sec-methodology}.}   The biggest difference is in the survival of \ion{H}{I} clouds, with $\tmin = 300$ K predicting longer lived clouds.  Still, the impact on \ion{H}{I} observations is quite small.  The ratios of the three observables is also quite constant, generally to within 30\%, as can be see from an inspection of the right two panels of Figures~\ref{fig:xrayproj}, \ref{fig:haproj}, and \ref{fig:hiproj}, and from Tables~\ref{tbl-xtail} and \ref{tbl-hatail}.

\subsection{Resolution}\label{sec-resolution}

Resolution is most likely to affect the survival and structure of our dense clouds in the tail.  The most direct results would be different amounts of neutral gas and H$\alpha$ emission.  A number of examinations of the survival of clouds with a variety of physics included have been performed.  Mellema et al. (2002) and Yirak et al. (2009) include radiative cooling.  Fragile et al. (2004) also discuss how including self-gravity increases the lifetime of clouds.  Nakamura et al. (2006) discuss the impact of using smooth cloud boundaries on the growth of instabilities.  They (and Yirak et al. 2009) find that a low density gradient results in slower growth of instabilities, which can retard cloud destruction.  Our most dense clouds have an analytically calculated destruction timescale due to turbulent viscous stripping (using eq. (22) in Nulsen 1982) of more than 1 Gyr, so resolution should not have a large impact over the timescales we consider in these simulations.  We use the same cloud parameters as in the evaporation time calculation, although even with an order of magnitude lower density clouds ($\sim$ 10$^{-24}$ g cm$^{-3}$), the destruction timescale is still about 200 Myr, which is much longer than the time at which we make our projections.  We use a relative velocity difference of 400 km s$^{-1}$ (although we do not show velocity plots in this paper, see TB10 for a detailed discussion of the tail velocity).  However, lower density clouds in our simulated tails cool and are compressed by the ICM, which may instead be destroyed by the ICM wind if we had better resolution that resulted in a steeper density gradient at the cloud edge. 

Although we do not perform a detailed resolution study of the runs in this paper, we do compare T3vl to a run with only 5 allowed levels of refinement, to a minimum cell size of 76 pc, a factor of two worse in resolution.  We compare outputs with the same amount of gas stripped from the galaxy.  The total X-ray luminosity of the lower resolution tail is $\sim$90\% of the luminosity of T3vl (i.e. only a 10\% change), while the total H$\alpha$ flux is only 28\% of T3vl.  The total HI flux is also decreased by a similar amount (to 33\% of the value in T3vl).  For both resolutions, there is a good correspondence between the relative HI and H$\alpha$ level and morphology; therefore, we argue that the ratio of HI to H$\alpha$ is more robust than the absolute level of either.

Both the small discrepancy in the X-ray luminosity and larger discrepancy in the H$\alpha$ flux can be explained by considering what gas is producing the emission.  X-ray emission is produced by gas that is mixed with the ICM and is not localized near the dense clouds in the tail, and so is not very dependent on high resolution.  One explanation for the slightly lower X-ray luminosity is that mixing happens more quickly and the gas is heated out of the X-ray bright regime.  

Unlike the diffuse nature of the X-ray emitting gas, H$\alpha$ emission occurs at the edges of clouds.  We find that in projection the range of H$\alpha$ intensities is the same as T3vl, meaning that the different resolution element size does not strongly affect the density and temperature at the edges of the dense clouds (and therefore does not strongly affect the H$\alpha$ emission). The main difference is that there are fewer dense clouds in the less-resolved wind.  In fact, the total \ion{H}{I} column density in the low resolution tail is about 30\% of the total \ion{H}{I} column density in T3vl--very similar to the H$\alpha$ flux fraction.  In projection, there are about half as many pixels with \ion{H}{I} column densities over 10$^{19}$ cm$^{-2}$.  Since we expect the total \ion{H}{I} column density to also depend on the number of clouds along the line of sight of the projection, we find that there should be roughly 26\% as many clouds (by simply squaring the 51\% we find in the projection plane).  Of course, clouds are more than a single pixel, so this is a rough approximation.  However, this indicates that the amount of H$\alpha$ emission closely follows the amount of \ion{H}{I} column density, and likely the number of \ion{H}{I} clouds.  There are fewer \ion{H}{I} clouds in the low resolution run because with lower resolution the maximum density is lower and therefore the clouds are destroyed and mixed more quickly into the ICM.  

Deep, high resolution observations in \ion{H}{I} will allow us to determine how many and for how long dense clouds survive, which will point to one of these mixing scenarios.  

\section{Conclusions}

We have run detailed galaxy simulations including radiative cooling and made comparisons to the observed tail of ESO 137-001 to understand the physical mechanisms at work in the ICM.  We compare three cases in which we vary our cooling floor between 8,000 K or 300 K and the ICM parameters as shown in Table \ref{tbl-runs}.  Our main conclusions are as follows:

1)  The X-ray luminosity of our simulated tails are an excellent match to the X-ray luminosity of the tail of ESO 137-001.  This suggests that we are correctly modeling the phase distribution of gas in the tail, and that the mixing of the hotter stripped ISM (T $>$ 10$^5$ K) is being accurately modeled in the simulations.

2)  We find that bright X-ray emission depends upon a high surrounding ICM pressure, and find a minimum necessary pressure of 9 $\times$ 10$^{-12}$ erg cm$^{-3}$.  This conclusion agrees well with the local environment in clusters where bright X-ray tails are observed.

3)  We compare our H$\alpha$ fluxes to the total diffuse H$\alpha$ flux measure by Sun et al. (2007).  As in our X-ray emission, we find an excellent match between our simulations and observations. 

4)  We predict that deeper observations will find \ion{H}{I} gas to at least 40 kpc above the disk, because diffuse H$\alpha$ emission directly traces neutral clouds (and has been observed to 40 kpc above the disk).  The observations of molecular hydrogen by Sivanandam et al. (2009) strengthen this prediction.  We also conclude that the mismatch between T3vl and Vollmer et al. (2001) indicates that the higher velocity cases better match the observations.

5)  Using the fact that H$\alpha$ traces neutral gas and calculating the evaporation times of our simulated clouds (based on their sizes and densities), we calculated a maximum efficiency for heat conduction of 24\% (T3vh and T3vl) or 12\% (T80vh).

By modeling a simulation using a high ICM pressure, we have shown that X-ray emission can coexist with \ion{H}{I} and H$\alpha$ emission.  We have also found that \ion{H}{I} and H$\alpha$ emission spatially coincide because H$\alpha$ is mostly produced at the edges of neutral clouds (Figure \ref{fig:slices}), while X-ray emission is generated in hot, diffuse gas that is mixing with the ICM.  This is seen by the more even distribution in the tail (Figures \ref{fig:xrayproj} and \ref{fig:slices}), and by the fact that the X-ray tail can be longer than the \ion{H}{I} and H$\alpha$ tails.  

Our excellent agreement with X-ray observations gives us confidence that we are correctly modeling the mixing, cooling and heating rate of X-ray emitting gas.  This in turn means that heat conduction is not acting strongly to heat the diffuse gas in the tail, and that small scale turbulence (below our resolution scale) is not quickly mixing stripped gas.  Both of these mechanisms would act to heat the gas in the tail out of the X-ray bright range and therefore lower the total luminosity, which would make our agreement to the observations worse.  We cannot rule out the possibility that we have less gas mass at higher emissivities than in ESO 137-001.  

We also find excellent agreement between our simulated H$\alpha$ flux and observations of diffuse H$\alpha$ emission.  We robustly conclude that diffuse H$\alpha$ emission coincides with \ion{H}{I} gas, and we find that H$\alpha$ emission outlines \ion{H}{I} clouds in all of our simulations.  The agreement between the total H$\alpha$ emission in our cases with different cooling floors underscores the fact that the edges of clouds, which are interacting with the ICM, emit the most strongly in H$\alpha$, not the central regions of the clouds, which are more likely to have radiatively cooled to the minimum allowed temperature.  Although our agreement with observations indicates that we may also be correctly modeling the edges of dense clouds, we cannot dismiss the possibility that we are not fully resolving the cloud edges.  In Section \ref{sec-resolution} we discuss how lower resolution results in fewer \ion{H}{I} clouds, and less H$\alpha$ emission.  Therefore, we cannot rule out that properties of our simulations that have not converged at our current resolution--the number of clouds, the rate of the decline of density at the edge of the cloud and the smallest scale of turbulent heating--have combined with the lack of heat conduction in our simulations to result in an accidental, and incorrect, agreement with observations.  Observations of \ion{H}{I} in the tail will provide an important check to these results. 

This work highlights the importance of comparing simulations to detailed, multi-wavelength observations of individual systems.  We are able to make predictions about this particular galaxy, such as the existence of \ion{H}{I} gas in the tail and that its three-dimensional velocity relative to the ICM is probably larger than 1413 km s$^{-1}$.  However, our simulation is not able to make any prediction about why there is a separated tail in ESO 137-001.  We also do not reproduce the exact surface brightness distribution along the X-ray tail.  As we have discussed, our goal was not to reproduce this specific tail.  In order to do this, we would recommend modeling a smaller galaxy and matching the inclination angle between the galaxy and the ICM wind. 

Using our comparison with the observations of ESO 137-001, we also draw more general conclusions about the importance of turbulence and the efficiency of heat conduction in the ICM.  We conclude that the mixing rate of the hot stripped ISM (T $>$ 10$^5$ K) is well-modeled in our simulations using only adiabatic compression and turbulent mixing down to the resolution of our simulations (38 pc).  We call upon observers to test our predictions of where in clusters tails will be X-ray bright, and to use deep observations to verify the connection between H$\alpha$ and \ion{H}{I} gas.  Observations of the \ion{H}{I} tail of ESO 137-001 can be used to test mixing of cold clouds into the ICM in our simulations.  

\acknowledgements

We acknowledge support from NSF grants AST-0547823, AST-0908390, and AST-1008134, as well as computational resources from the National Center for Supercomputing Applications, NSF Teragrid, and Columbia University's Hotfoot cluster.  We thank Jacqueline van Gorkom, Jeffrey Kenney and the referee Bernd Vollmer for useful discussions, as well as Elizabeth Tasker for invaluable help setting up the initial conditions.


\begin{thebibliography}

\bibitem[Arnaud(1996)]{1996ASPC..101...17A} Arnaud, K.~A.\ 1996, 
Astronomical Data Analysis Software and Systems V, 101, 17 

\bibitem[Boehringer et al.(1996)]{1996ApJ...467..168B} Boehringer, H., 
Neumann, D.~M., Schindler, S., 
\& Kraan-Korteweg, R.~C.\ 1996, \apj, 467, 168 


\bibitem[Briel et 
al.(1992)]{1992A&A...259L..31B} Briel, U.~G., Henry, J.~P., \& Boehringer, H.\ 1992, \aap, 259, L31 


\bibitem[Bryan(1999)]{1999CoScE...1...46B} Bryan, G.~L.\ 1999, 
Comput.~Sci.~Eng., Vol.~1, No.~2, p.~46 - 53, 1, 46 


\bibitem[Burkert(1995)]{1995ApJ...447L..25B} Burkert, A.\ 1995, \apjl, 447, 
L25 

\bibitem[Cavagnolo et al.(2009)]{2009ApJS..182...12C} Cavagnolo, K.~W., 
Donahue, M., Voit, G.~M., \& Sun, M.\ 2009, \apjs, 182, 12 



\bibitem[Chung et al.(2007)]{2007ApJ...659L.115C} Chung, A., van Gorkom, 
J.~H., Kenney, J.~D.~P., \& Vollmer, B.\ 2007, \apjl, 659, L115 


\bibitem[Cowie 
\& McKee(1977)]{1977ApJ...211..135C} Cowie, L.~L., \& McKee, C.~F.\ 1977, \apj, 211, 135 


\bibitem[Donnelly et al.(1998)]{1998ApJ...500..138D} Donnelly, R.~H., 
Markevitch, M., Forman, W., Jones, C., David, L.~P., Churazov, E., 
\& Gilfanov, M.\ 1998, \apj, 500, 138 

\bibitem[Ettori et al.(1998)]{1998MNRAS.300..837E} Ettori, S., Fabian, 
A.~C., \& White, D.~A.\ 1998, \mnras, 300, 837 


\bibitem[Ferland et al.(1998)]{1998PASP..110..761F} Ferland, G.~J., 
Korista, K.~T., Verner, D.~A., Ferguson, J.~W., Kingdon, J.~B., 
\& Verner, E.~M.\ 1998, \pasp, 110, 761 


\bibitem[Fragile et al.(2004)]{2004ApJ...604...74F} Fragile, P.~C., Murray, 
S.~D., Anninos, P., \& van Breugel, W.\ 2004, \apj, 604, 74 


\bibitem[Gnedin(2003)]{2003ApJ...589..752G} Gnedin, O.~Y.\ 2003, \apj, 589, 
752 


\bibitem[Haardt 
\& Madau(2001)]{2001cghr.confE..64H} Haardt, F., \& Madau, P.\ 2001, Clusters of Galaxies and the High Redshift Universe Observed in X-rays,  


\bibitem[Haynes et al.(2007)]{2007ApJ...665L..19H} Haynes, M.~P., 
Giovanelli, R., \& Kent, B.~R.\ 2007, \apjl, 665, L19 


\bibitem[Hernquist(1993)]{1993ApJS...86..389H} Hernquist, L.\ 1993, \apjs, 
86, 389 


\bibitem[Irwin 
\& Sarazin(1996)]{1996ApJ...471..683I} Irwin, J.~A., \& Sarazin, C.~L.\ 1996, \apj, 471, 683 


\bibitem[Kapferer et 
al.(2009)]{2009A&A...499...87K} Kapferer, W., Sluka, C., Schindler, S., Ferrari, C., \& Ziegler, B.\ 2009, \aap, 499, 87 


\bibitem[Kenney et al.(2008)]{2008ApJ...687L..69K} Kenney, J.~D.~P., Tal, 
T., Crowl, H.~H., Feldmeier, J., \& Jacoby, G.~H.\ 2008, \apjl, 687, L69 


\bibitem[Kim et al.(2008)]{2008ApJ...688..931K} Kim, D.-W., Kim, E., 
Fabbiano, G., \& Trinchieri, G.\ 2008, \apj, 688, 931 


\bibitem[Koopmann et al.(2008)]{2008ApJ...682L..85K} Koopmann, R.~A., et 
al.\ 2008, \apjl, 682, L85 


\bibitem[Kronberger et 
al.(2008)]{2008A&A...481..337K} Kronberger, T., Kapferer, W., Ferrari, C., Unterguggenberger, S., \& Schindler, S.\ 2008, \aap, 481, 337 


\bibitem[Machacek et al.(2006)]{2006ApJ...644..155M} Machacek, M., Jones, 
C., Forman, W.~R., \& Nulsen, P.\ 2006, \apj, 644, 155 


\bibitem[Mellema et 
al.(2002)]{2002A&A...395L..13M} Mellema, G., Kurk, J.~D., {\ Rouml}ttgering, H.~J.~A.\ 2002, \aap, 395, L13 


\bibitem[Miyamoto 
\& Nagai(1975)]{1975PASJ...27..533M} Miyamoto, M., \& Nagai, R.\ 1975, \pasj, 27, 533 

\bibitem[Mohr et al.(1999)]{1999ApJ...517..627M} Mohr, J.~J., Mathiesen, 
B., \& Evrard, A.~E.\ 1999, \apj, 517, 627 


\bibitem[Mori 
\& Burkert(2000)]{2000ApJ...538..559M} Mori, M., \& Burkert, A.\ 2000, \apj, 538, 559 


\bibitem[Norman 
\& Bryan(1999)]{1999ASSL..240...19N} Norman, M.~L., \& Bryan, G.~L.\ 1999, Numerical Astrophysics, 240, 19 


\bibitem[Nulsen(1982)]{1982MNRAS.198.1007N} Nulsen, P.~E.~J.\ 1982, \mnras, 
198, 1007 


\bibitem[O'Shea et al.(2004)]{2004astro.ph..3044O} O'Shea, B.~W., Bryan, 
G., Bordner, J., Norman, M.~L., Abel, T., Harkness, R., 
\& Kritsuk, A.\ 2004, arXiv:astro-ph/0403044 


\bibitem[Oosterloo 
\& van Gorkom(2005)]{2005A&A...437L..19O} Oosterloo, T., \& van Gorkom, J.\ 2005, \aap, 437, L19 


\bibitem[Quilis et al.(2000)]{2000Sci...288.1617Q} Quilis, V., Moore, B., 
\& Bower, R.\ 2000, Science, 288, 1617 


\bibitem[Raymond 
\& Smith(1977)]{1977ApJS...35..419R} Raymond, J.~C., \& Smith, B.~W.\ 1977, \apjs, 35, 419 

\bibitem[Roediger et al.(2006)]{2006MNRAS.371..609R} Roediger, E., 
Br{\"u}ggen, M., \& Hoeft, M.\ 2006, \mnras, 371, 609

\bibitem[Roediger 
\& Br{\"u}ggen(2008)]{2008MNRAS.388..465R} Roediger, E., \& Br{\"u}ggen, M.\ 2008, \mnras, 388, 465 


\bibitem[Schulz 
\& Struck(2001)]{2001MNRAS.328..185S} Schulz, S., \& Struck, C.\ 2001, \mnras, 328, 185 


\bibitem[Sivanandam et al.(2009)]{2009arXiv0912.0075S} Sivanandam, S., 
Rieke, M.~J., \& Rieke, G.~H.\ 2009, arXiv:0912.0075 

\bibitem[Skrutskie et al.(2006)]{2006AJ....131.1163S} Skrutskie, M.~F., et 
al.\ 2006, \aj, 131, 1163 

\bibitem[Sun et al.(2010)]{2010ApJ...708..946S} Sun, M., Donahue, M., 
Roediger, E., Nulsen, P.~E.~J., Voit, G.~M., Sarazin, C., Forman, W., 
\& Jones, C.\ 2010, \apj, 708, 946 


\bibitem[Sun et al.(2007)]{2007ApJ...671..190S} Sun, M., Donahue, M., 
\& Voit, G.~M.\ 2007, \apj, 671, 190 


\bibitem[Sun et al.(2006)]{2006ApJ...637L..81S} Sun, M., Jones, C., Forman, 
W., Nulsen, P.~E.~J., Donahue, M., \& Voit, G.~M.\ 2006, \apjl, 637, L81 


\bibitem[Sun 
\& Vikhlinin(2005)]{2005ApJ...621..718S} Sun, M., \& Vikhlinin, A.\ 2005, \apj, 621, 718 


\bibitem[Tasker 
\& Bryan(2006)]{2006ApJ...641..878T} Tasker, E.~J., \& Bryan, G.~L.\ 2006, \apj, 641, 878 


\bibitem[Tonnesen 
\& Bryan(2010)]{2010ApJ...709.1203T} Tonnesen, S., \& Bryan, G.~L.\ 2010, \apj, 709, 1203 


\bibitem[Tonnesen 
\& Bryan(2009)]{2009ApJ...694..789T} Tonnesen, S., \& Bryan, G.~L.\ 2009, \apj, 694, 789 


\bibitem[Trachternach et al.(2008)]{2008AJ....136.2720T} Trachternach, C., 
de Blok, W.~J.~G., Walter, F., Brinks, E., 
\& Kennicutt, R.~C.\ 2008, \aj, 136, 2720 


\bibitem[Vollmer et 
al.(2005)]{2005A&A...441..473V} Vollmer, B., Braine, J., Combes, F., \& Sofue, Y.\ 2005, \aap, 441, 473 


\bibitem[Vollmer et al.(2001)]{2001ApJ...561..708V} Vollmer, B., Cayatte, 
V., Balkowski, C., \& Duschl, W.~J.\ 2001, \apj, 561, 708 


\bibitem[Vollmer et 
al.(2001)]{2001A&A...369..432V} Vollmer, B., Cayatte, V., van Driel, W., Henning, P.~A., Kraan-Korteweg, R.~C., Balkowski, C., Woudt, P.~A., \& Duschl, W.~J.\ 2001, \aap, 369, 432 


\bibitem[Vollmer 
\& Huchtmeier(2007)]{2007A&A...462...93V} Vollmer, B., \& Huchtmeier, W.\ 2007, \aap, 462, 93 


\bibitem[Vollmer et 
al.(2008)]{2008A&A...483...89V} Vollmer, B., Soida, M., Chung, A., van Gorkom, J.~H., Otmianowska-Mazur, K., Beck, R., Urbanik, M., \& Kenney, J.~D.~P.\ 2008, \aap, 483, 89 


\bibitem[Vollmer et 
al.(2006)]{2006A&A...453..883V} Vollmer, B., Soida, M., Otmianowska-Mazur, K., Kenney, J.~D.~P., van Gorkom, J.~H., \& Beck, R.\ 2006, \aap, 453, 883 


\bibitem[Wang et al.(2004)]{2004ApJ...611..821W} Wang, Q.~D., Owen, F., 
\& Ledlow, M.\ 2004, \apj, 611, 821 


\bibitem[Woudt et 
al.(2004)]{2004A&A...415....9W} Woudt, P.~A., Kraan-Korteweg, R.~C., Cayatte, V., Balkowski, C., \& Felenbok, P.\ 2004, \aap, 415, 9 


\bibitem[Woudt et 
al.(1998)]{1998A&A...338....8W} Woudt, P.~A., Kraan-Korteweg, R.~C., Fairall, A.~P., Boehringer, H., Cayatte, V., \& Glass, I.~S.\ 1998, \aap, 338, 8 


\bibitem[Woudt et al.(2008)]{2008MNRAS.383..445W} Woudt, P.~A., 
Kraan-Korteweg, R.~C., Lucey, J., Fairall, A.~P., 
\& Moore, S.~A.~W.\ 2008, \mnras, 383, 445 


\bibitem[Yagi et al.(2007)]{2007ApJ...660.1209Y} Yagi, M., Komiyama, Y., 
Yoshida, M., Furusawa, H., Kashikawa, N., Koyama, Y., 
\& Okamura, S.\ 2007, \apj, 660, 1209 


\bibitem[Yirak et al.(2009)]{2009arXiv0912.4777Y} Yirak, K., Frank, A., 
\& Cunningham, A.~J.\ 2009, arXiv:0912.4777 


\bibitem[Yoshida et al.(2004)]{2004AJ....127.3653Y} Yoshida, M., et al.\ 
2004, \aj, 127, 3653 


\bibitem[Yoshida et al.(2004)]{2004AJ....127...90Y} Yoshida, M., et al.\ 
2004, \aj, 127, 90 

\bibitem[Yoshida et al.(2008)]{2008ApJ...688..918Y} Yoshida, M., et al.\ 
2008, \apj, 688, 918 

\bibitem[Yoshida et al.(2002)]{2002ApJ...567..118Y} Yoshida, M., et al.\ 
2002, \apj, 567, 118 

\bibitem[Zhang et 
al.(2005)]{2005A&A...429...85Z} Zhang, Y.-Y., B{\"o}hringer, H., Mellier, Y., Soucail, G., \& Forman, W.\ 2005, \aap, 429, 85 

\end{thebibliography}
\end{document}